\documentclass[%
 reprint,
 amsmath,amssymb,
 aps,prl,
]{revtex4-1}
\usepackage{float} 
\usepackage{color}
\usepackage{graphicx}
\usepackage{dcolumn}
\usepackage{bm}
\usepackage{hyperref}
\usepackage{footnote}

\newcommand{\rp}{\mathbf{r}}
\newcommand{\np}{\mathbf{\hat{n}}}
\newcommand{\bp}{\mathbf{b}}
\DeclareMathOperator*{\argmin}{arg\,min}
\DeclareMathOperator*{\argmax}{arg\,max}

\begin{document}

\preprint{APS/123-QED}

\title{Three-body Hydrogen Bond Defects Contribute Significantly to the Dielectric Properties of the Liquid Water-Vapor Interface}

\author{Sucheol Shin}
\author{Adam P. Willard}
\email{awillard@mit.edu}
\affiliation{ 
Department of Chemistry, Massachusetts Institute of Technology, Cambridge, Massachusetts 02139, USA
}

\date{\today}

\begin{abstract}
In this Letter, we present a simple model of aqueous interfacial molecular structure and we use this model to isolate the effects of hydrogen bonding on the dielectric properties of the liquid water-vapor interface. 
By comparing this model to the results of atomistic simulation we show that the anisotropic distribution of molecular orientations at the interface can be understood by considering the behavior of a single water molecule interacting with the average interfacial density field via an empirical hydrogen bonding potential. 
We illustrate that the depth dependence of this orientational anisotropy is determined by the geometric constraints of hydrogen bonding and we show that the primary features of simulated orientational distributions can be reproduced by assuming an idealized, perfectly tetrahedral hydrogen bonding geometry. 
We also demonstrate that non-ideal hydrogen bond geometries are required to produce interfacial variations in the average orientational polarization and polarizability. 
We find that these interfacial properties contain significant contributions from a specific type of geometrically distorted three-body hydrogen bond defect that is preferentially stabilized at the interface. 
Our findings thus reveal that the dielectric properties of the liquid water-vapor interface are determined by collective molecular interactions that are unique to the interfacial environment. 

\end{abstract}

\pacs{Valid PACS appear here}

\maketitle

The dielectric properties of liquid water are determined in large part by the orientational fluctuations of dipolar water molecules \cite{Kirkwood1939,Frohlich1958,Bopp1996,Despa2004}.
Near a liquid water-vapor interface these orientational fluctuations are anisotropic, leading to dielectric properties that differ significantly from their bulk values \cite{Bonthuis2011,Vacha2012,Schlaich2016a}.
These differences are fundamental to interface-selective chemical and physical processes \cite{Saykally2013, Jungwirth2006, Levin2009, Zhao2011}, but they are generally difficult to 
access experimentally.
In this Letter, we study these differences with a statistical mechanical model of interfacial hydrogen bonding.
This model provides the ability to relate the microscopic characteristics of hydrogen bonding to the emergent properties of the liquid water interface.
By comparing this model to the results of atomistic simulation we isolate the specific role of hydrogen bonding in determining interfacial molecular structure.
We demonstrate that water's interfacial molecular structure is determined by the interplay between the anisotropic features of the interfacial density field and the molecular geometry of hydrogen bonding interactions.
While many details of interfacial molecular structure reflect an idealized tetrahedral hydrogen bonding geometry, we show that only distorted non-tetrahedral hydrogen bond structures can contribute to changes in interfacial dielectric properties. 
We identify a specific type of non-tetrahedral hydrogen bond defect that is preferentially stabilized at the interface and show that it contributes significantly to the unique dielectric properties of the interfacial environment.

Our microscopic understanding of interfacial molecular structure derives primarily from a combination of surface-sensitive experiments, such as vibrational sum frequency generation \cite{Ji2008, Shen2006, Raymond2002, Raymond2003}, and atomistic simulation~\cite{Wilson1987, Matsumoto1988, Fan2009, Willard2010}. 
This combination has revealed that the interfacial environment contains depth dependent molecular populations that vary in their orientational alignments.  
The details of this depth-dependent orientational molecular structure cannot be determined from existing experimental data alone and thus we rely on molecular simulation to supplement our molecular level understanding.
The molecular structure predicted through the use of standard classical force fields, such as SPC/E~\cite{Berendsen1987} and TIP5P~\cite{Mahoney2000}, are widely used in the study of water interfaces because they are both computationally efficient and have been shown to be consistent with available experimental data~\cite{Pieniazek2011, Nihonyanagi2011}.

Even at microscopic length scales the position of a liquid-vapor interface exhibits capillary wave-like spatial undulations~\cite{Ismail2006, Sedlmeier2009}.
These undulations contribute to the properties of the interface but they also serve to blur out molecular scale details.
The microscopic details of the actual phase boundary, \textit{i.e.}, the \textit{intrinsic interface} is difficult to isolate experimentally.
In theoretical studies, however, the intrinsic interface and its spatial deformations can be treated separately. 
This Letter focuses on the intrinsic interface, defining all molecular coordinates relative to the time varying position of the instantaneous liquid phase boundary~\cite{Willard2010}.
We identify the position of the phase boundary following the procedure described in Ref.~\onlinecite{Willard2010}.
Our model is thus most valid over microscopic length scales, where the effects of surface roughening are negligible, however, since the deformations in the liquid-vapor phase boundary are well characterized by capillary wave theory \cite{Buff1965, Mecke1999, Rowlinson2002, Patel2011a}, their effects are trivial to reincorporate to enable comparison to experiment.

Atomistic simulations have revealed that the molecular structure of the intrinsic water-vapor interface is anisotropic extending about 1 nm into the bulk liquid \cite{Willard2010, Willard2014}. 
This anisotropic molecular structure can be decomposed into two separate components: 
(1) an interfacial density field, $\rho(\rp)$, where $\rp$ denotes the position measured relative to the instantaneous interface, and 
(2) a position-dependent probability distribution for molecular orientations, $P(\vec{\kappa} \vert \rp)$, where $\vec{\kappa}$ uniquely specifies the rotational configuration of a water molecule.
The interplay between these two components is mediated by a combination of molecular packing effects and collective hydrogen bonding interactions. 
Classical density functional theory can be used to compute $\rho(\rp)$, but it does not provide explicit information about $P(\vec{\kappa} \vert \rp)$~\cite{Hughes2013, Rowlinson2002}.
Here we present a theoretical model for computing $P(\vec{\kappa} \vert \rp)$ from $\rho(\rp)$, noting that the complete intrinsic interfacial molecular structure can therefore be derived by combining this model with an existing approach for computing $\rho(\rp)$.

\begin{figure}[h]
\centering
\includegraphics[width = 3.4 in]{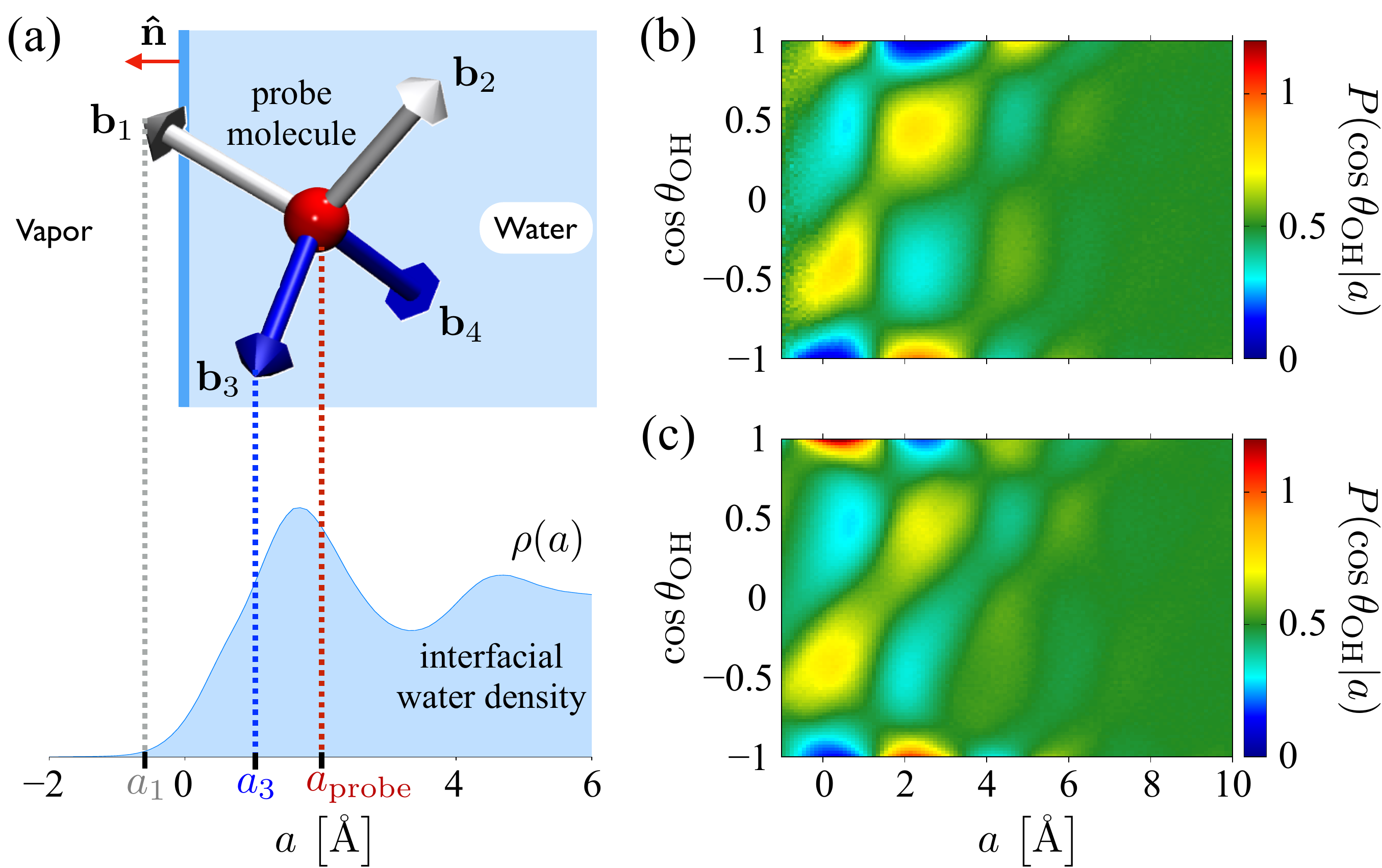}
\caption{(a) Schematic depiction of the mean-field model showing a probe molecule with tetrahedrally coordinated bond vectors (white for donor, blue for acceptor) within the liquid (blue shaded region) at a distance $a_\text{probe}$ from the position of the instantaneous interface (solid blue line).
A plot of the interfacial density profile, $\rho(a)$, obtained from the MD simulation with TIP5P water \cite{SM}, is shown with dotted lines indicating the termination points of bond vectors $\bp_1$ and $\bp_3$. 
Panels (b) and (c) contain plots of the orientational distribution function, $P(\cos \theta_\mathrm{OH} \vert a)$ (see Eq.~(\ref{eq:red_dist})), as indicated by shading, computed from atomistic simulation and from the rigid tetrahedral model respectively.}
\label{fig:1}
\end{figure}

Our model utilizes a mean field approach by computing $P(\vec{\kappa} \vert \rp)$ based on the orientational preferences of a single probe molecule immersed in the anisotropic mean density field of the intrinsic interface, $\rho(\rp)$. 
As illustrated in Fig.~\ref{fig:1}(a), the probe molecule is modeled as a point particle with four tetrahedrally coordinated hydrogen bond vectors, denoted $\bp_1$, $\bp_2$, $\bp_3$, and $\bp_4$.
The length of these vectors are chosen to correspond to that of a hydrogen bond, so that each vector indicates the preferred position of a hydrogen bond partner. 
To mimic the hydrogen bonding properties of water, each bond vector is assigned a directionality, with $\bp_1$ and $\bp_2$ acting as hydrogen bond donors and $\bp_3$ and $\bp_4$ acting as hydrogen bond acceptors.  
The absolute orientations of these tetrahedral hydrogen bond vectors are specified by $\vec{\kappa}$.

The probe molecule interacts with the interfacial density field via an empirical hydrogen bonding potential,
$E(\vec{\kappa},\rp,\lbrace n_k \rbrace)$, which specifies the potential energy of a probe molecule with position $\rp$, orientation $\vec{\kappa}$, and hydrogen bonding configuration $\lbrace n_k \rbrace$.
We define $\lbrace n_k \rbrace \equiv (n_1,n_2,n_3,n_4)$, where $n_i$ is a binary variable that indicates the hydrogen bonding state of the $i$th bond vector.
Specifically, $n_i=1$ if the probe molecule has formed a hydrogen bond along $\bp_i$, and $n_i=0$ if it has not.
There are many possible ways to define this empirical hydrogen bonding potential.
We begin by considering the simple form, 
\begin{equation}
E(\vec{\kappa}, \rp, \lbrace n_k \rbrace) = \sum_{i = 1}^{4} \epsilon_\text{w} n_i(\rp, \bp_i)~,
\label{eq:energy1}
\end{equation}
where $\epsilon_\text{w}$ is the energy associated with forming a hydrogen bond.
We treat the $n_i$'s as independent random variables that are distributed according to,
\begin{equation}
	n_i(\rp,\bp_i)= \left\{ \begin{array}{ll}
		1, & \quad   \textrm{with probability } P_\mathrm{HB}(\rp_i), \\
          	0, & \quad \textrm{with probability } 1-P_\mathrm{HB}(\rp_i), \end{array} \right.
	\label{eq:roa_ni}
\end{equation}
where $\rp_i=\rp+\bp_i$ denotes the terminal position of the $i$th bond vector and $P_\mathrm{HB}(\rp_i)$ specifies the probability for successful hydrogen bonding at position $\rp_i$.

In the context of this model, the probability for a molecule at position $\rp$ to adopt an orientation, $\vec{\kappa}$, can thus be expressed as,
\begin{equation}
P(\vec{\kappa} \vert \rp) = \left \langle e^{-\beta E(\vec{\kappa},\rp,\lbrace n_k \rbrace)} \right \rangle_\mathrm{b}/Z(\rp),
\label{eq:dist1}
\end{equation}
where $\langle \cdots \rangle_\mathrm{b}$ denotes an average over all possible hydrogen bonding states (\textit{i.e.}, variations in the $n_i$'s), $1/\beta$ is the Boltzmann constant, $k_{B}$, times temperature $T$, and $Z(\rp)= \int d\vec{\kappa} \left \langle e^{-\beta E(\vec{\kappa},\rp,\lbrace n_k \rbrace)} \right \rangle_\mathrm{b}$ is the orientational partition function for the probe molecule at position $\rp$.
By evaluating the average explicitly, the numerator of Eq.~(\ref{eq:dist1}) can be written as,
\begin{equation}
\left \langle e^{-\beta E(\vec{\kappa},\rp, \lbrace n_k \rbrace)} \right \rangle_\mathrm{b} = \prod_{i=1}^4 \left [ 1 + P_\mathrm{HB} \left (\rp_i)(e^{-\beta \epsilon_\text{w}} - 1 \right ) \right ],
\label{eq:dist2}
\end{equation}
and this equation, when combined with Eq.~(\ref{eq:dist1}) provides a general analytical framework for computing the orientational molecular structure of an interface with a given density profile, $\rho(\rp)$.

We simplify this general theoretical model by assuming that $P_\mathrm{HB}(\rp_i) \propto \rho(\rp_i)$, where the value of the proportionality constant is chosen to match bulk hydrogen bonding statistics.
Furthermore, we assume that the intrinsic interface is laterally isotropic and planar over molecular length scales.
According to this assumptions, functions that depend on $\rp$ to be expressed in terms of $a$, which denotes the scalar distance from the instantaneous interface measured perpendicular to the interfacial plane.
For instance, with this assumption the function $P_\mathrm{HB}(\rp)$ is simplified to $P_\mathrm{HB}(a_i) \propto \rho(a_i)$, where $a_i$ is given by $a_i = a - \bp_i \cdot \np$, and $\np$ is the unit vector normal to the plane of the interface (see Fig.~\ref{fig:1}(a)).

We parameterize our model by comparing to the results of atomistic simulations \cite{[See Supplemental Material at \protect{[}URL will be inserted by publisher\protect{]} for the computational details about atomistic simulations and interfacial structures of different classical water force fields\protect{,} optimization for the hydrogen bond energy parameter\protect{,} quantifying distortions in hydrogen bond geometry and associated energetic properties\protect{,} implementation of the three-body fluctuation model\protect{,} and application of mean-field models to the SPC/E force field]SM}. 
We perform this comparison using the reduced orientational distribution function,
\begin{equation}
P(\cos \theta_\mathrm{OH} \vert a) = \int d\vec{\kappa} P(\vec{\kappa}\vert a) \left [\frac{1}{2} \sum_{i=1}^2 \delta(\cos \theta_i - \cos \theta_\mathrm{OH}) \right ],
\label{eq:red_dist}
\end{equation}
where the summation is taken over the two donor bond vectors, $\cos \theta_i = \bp_i \cdot \np / \vert \bp_i \vert$, and $\delta (x)$ is the Dirac delta function.
Specifically, we determine free parameter, $\epsilon_\text{w}$, by minimizing the Kullback-Leibler divergence for $P(\cos \theta_\mathrm{OH} \vert a)$ computed from our model and that from atomistic simulation \cite{SM}. 
For the TIP5P force field~\cite{Mahoney2000} this parameterization yields $\epsilon_\text{w}=-3.47 \,\mathrm{kJ}/\mathrm{mol}$ at $T = 298\,\text{K}$.
We find that different water models yield similar value of $\epsilon_\mathrm{w}$ \cite{SM}.
We note that this value includes the effects of environmental stabilization on broken hydrogen bonds and is therefore significantly lower in magnitude than absolute hydrogen bond energies~\cite{Suresh2000,Jorgensen1983}.
This demonstrates that our parameterize hydrogen bond strength is similar to estimates based on X-ray absorption measurements~\cite{Smith2004, Smith2005}, which highlights that this effective bonding energy determines the relevant energy scale for fluctuations in the structure of aqueous hydrogen bond networks.

The effect of molecular dipole orientations on interfacial polarization and polarizability can be computed from $P(\vec{\kappa} \vert a)$. 
We specify orientational polarization in terms of the average dipole field, 
\begin{equation}
\langle \mu_{\np} (a) \rangle = \mu_\text{w}\int d\vec{\kappa} P(\vec{\kappa}\vert a) \left[\bm{\hat{\mu}} (\vec{\kappa}) \cdot \np\right] ,
\label{eq:mu_ave}
\end{equation}
where $\bm{\hat{\mu}}(\vec{\kappa})$ is the unit dipole vector of tagged molecule in particular orientation (\emph{i.e.}, $\bm{\hat{\mu}} = (\bp_1 + \bp_2)/\vert \bp_1 + \bp_2 \vert$) and $\mu_\mathrm{w}$ is the dipole moment of an individual water molecule.
Similarly, the orientational polarizability can be related to the fluctuations in the dipole field \cite{Frohlich1958},
\begin{equation}
\langle (\delta \mu_{\np}(a))^2 \rangle =\mu_\text{w}^2 \int d\vec{\kappa} P(\vec{\kappa}\vert a) \left[ \bm{\hat{\mu}} (\vec{\kappa}) \cdot \np \right]^2 - \langle \mu_{\np} (a) \rangle^2.
\label{eq:mu_fluct}
\end{equation}
We derive physical insight into the role of hydrogen bonding in water's interfacial molecular structure by comparing the functions in Eqs.~(\ref{eq:red_dist})-(\ref{eq:mu_fluct}) to the same quantities computed from atomistic simulations.

We first consider an idealized variation of our model in which the hydrogen bond vectors are of fixed length, $d_\mathrm{HB}=2.8 \,\mathrm{\AA}$ (\emph{i.e.}, the equilibrium hydrogen bond distance \cite{Jorgensen1983}), and rigidly arranged with an ideal tetrahedral geometry.
We refer to this model as the \textit{rigid tetrahedral model}.
In Fig.~\ref{fig:1}(b)-(c), we show that $P(\cos \theta_\mathrm{OH} \vert a)$ computed from the rigid tetrahedral model is similar to that computed from simulations with TIP5P water.
This similarity illustrates that the seemingly complicated depth dependent orientational patterns in $P(\cos \theta_\mathrm{OH} \vert a)$ have simple physical origins.
Namely, these patterns are determined by the constraints imposed on tetrahedral coordination by the anisotropic interfacial density field.

\begin{figure}[h]
\centering
\includegraphics[width = 3.4 in]{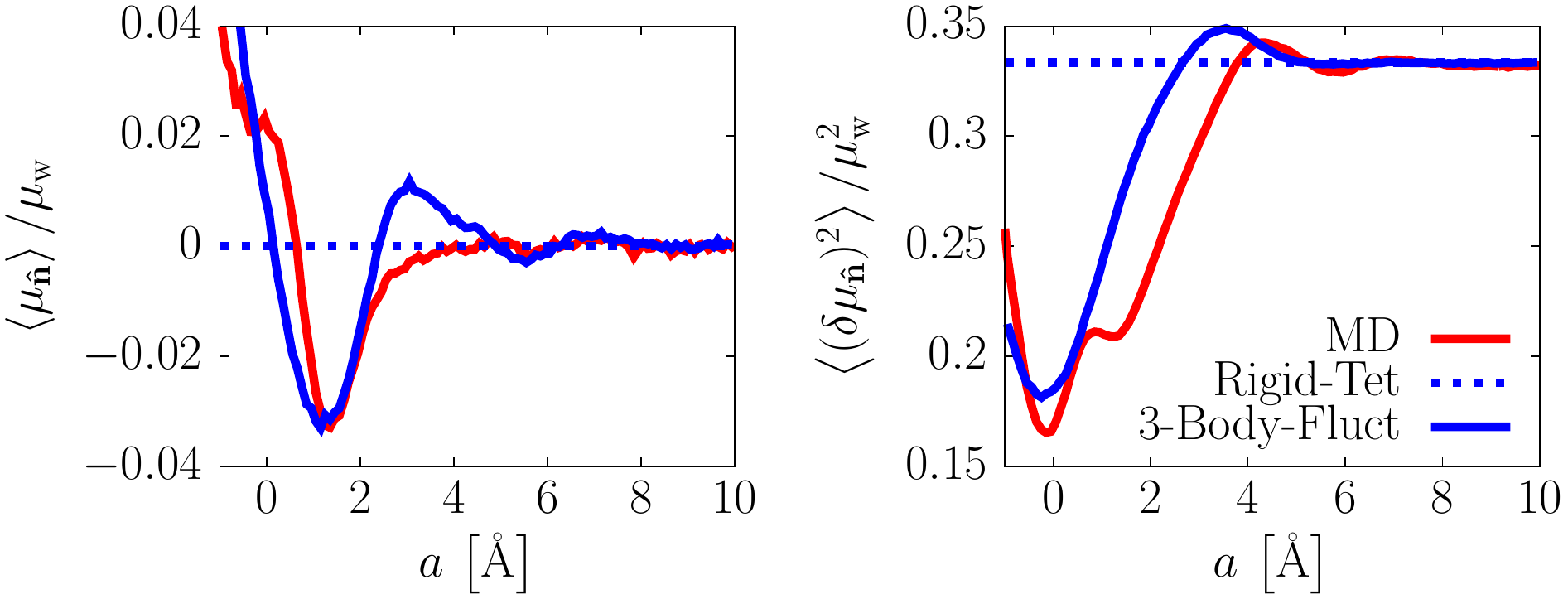}
\caption{Interfacial mean dipole orientation, $\langle \mu_{\np}(a) \rangle$, and dipole fluctuations, $\langle (\delta \mu_{\np}(a) )^2 \rangle$, computed from molecular dynamics (MD) simulation and different variations of our mean field model. 
Solid red lines correspond to atomistic simulation data \cite{SM}. Dashed (Rigid-Tet) and solid (3-Body-Fluct) blue lines correspond to the rigid tetrahedral model and the three-body fluctuation model (see Eq.~(\ref{eq:energy2})), respectively.}
\label{fig:2}
\end{figure}

As the atomistic simulation data in Fig.~\ref{fig:2} illustrates, both $\langle \mu_{\np}(a)\rangle$ and $\langle (\delta \mu_{\np}(a))^2 \rangle$ vary significantly from their bulk values at the liquid vapor interface.
However, these variations are not captured by the rigid tetrahedral model, despite the ability of this model to capture the primary features of $P(\cos \theta_\mathrm{OH} \vert a)$.
In fact, the rigid tetrahedral model predicts that $\langle \mu_{\np}(a)\rangle$ and $\langle (\delta \mu_{\np}(a))^2 \rangle$ are independent of depth (dashed linesd in Fig.~\ref{fig:2}).
This depth-independent behavior is a mathematical consequence of modeling hydrogen bonds as perfectly tetrahedral and energetically symmetric for donor and acceptor bonds
\footnote{
For the perfectly tetrahedral coordination, $\sum_{i=1}^{4} \bp_i = 0$. 
Because of the symmetry, any pair of distinct $i$ and $j$ can represent the donor bonds such that \unexpanded{$\langle \bm{\mu} \rangle = \mu_\text{w}\langle \bm{\hat{\mu}} \rangle \propto \langle \bp_i + \bp_j \rangle = 0$} for any orientational distribution. 
The dipole vector also can be decomposed such that \unexpanded{$\bm{\mu} = \sum_{k=1}^{3} \mu_{\mathbf{\hat{e}}_k} \mathbf{\hat{e}}_k$} where $\mathbf{\hat{e}}_k$'s are orthonormal vectors in three dimensions. Choosing $\mathbf{\hat{e}}_1 = \np$, we have \unexpanded{$\langle \mu_{\np} \rangle = \langle \bm{\mu} \cdot \np \rangle  = \langle \bm{\mu} \rangle \cdot \np = 0$}.
Here $\np$ could be other two unit vectors and there is no preference along specific direction. That is, since $|\bm{\mu}|^2 = \sum_{k=1}^{3} \mu_{\mathbf{\hat{e}}_k}^2 = \mu_\text{w}^2$, we expect \unexpanded{$\langle \mu_\np^2 \rangle = \mu_\text{w}^2/3$}.
}.
Interfacial variations in orientational polarization and polarizability must therefore arise through a combination of
(1) distortions in tetrahedral coordination geometry and
(2) asymmetry in donor/acceptor hydrogen bond energies.
We can study the specific influence of these effects on $\langle \mu_{\np}(a)\rangle$ and $\langle (\delta \mu_{\np}(a))^2 \rangle$ by explicitly including their effects in the empirical hydrogen bonding interaction of our mean field model. 

\begin{figure}[h]
\centering
\includegraphics[width = 3.4 in]{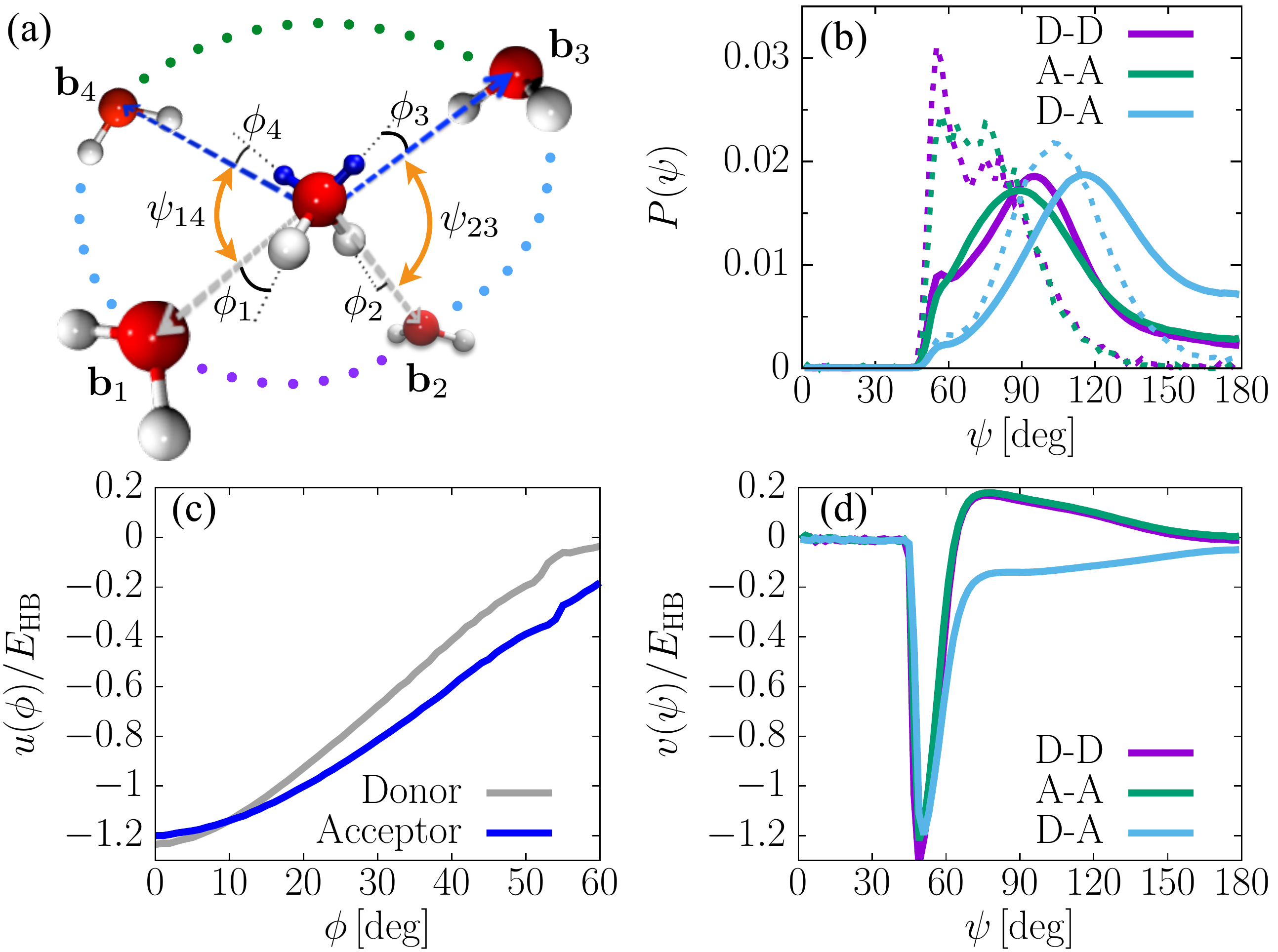}
\caption{(a) A schematic illustration of the angles used to quantify hydrogen bond geometries.
(b) Probability distributions for inter-bond angles, $\psi$, generated from atomistic simulation, computed separately for donor-donor (D-D), acceptor-acceptor (A-A), and donor-acceptor (D-A) pairs of bonds \cite{SM}.
Solid and dashed lines correspond to statistics generated within the bulk liquid and at the interface (\emph{i.e.}, $|a| < .1 \,\mathrm{\AA}$) respectively.
(c) Average direct interaction energy, $u(\phi)$, expressed in units of the average bulk hydrogen bond energy, $E_\mathrm{HB} = 20.9 \, \mathrm{kJ}/\mathrm{mol}$, between a tagged molecule and individual hydrogen bond partners, computed separately for donor and acceptor bonds.
(d) Average direct interaction energy, $v(\psi)$, between two hydrogen bond partners of a tagged molecule, as indicated by the dotted arcs in panel (a), computed separately for the case of two donor partners, two acceptor partners, and one acceptor and one donor.
}
\label{fig:3}
\end{figure}

\begin{figure}[h]
\centering
\includegraphics[width = 3.4 in]{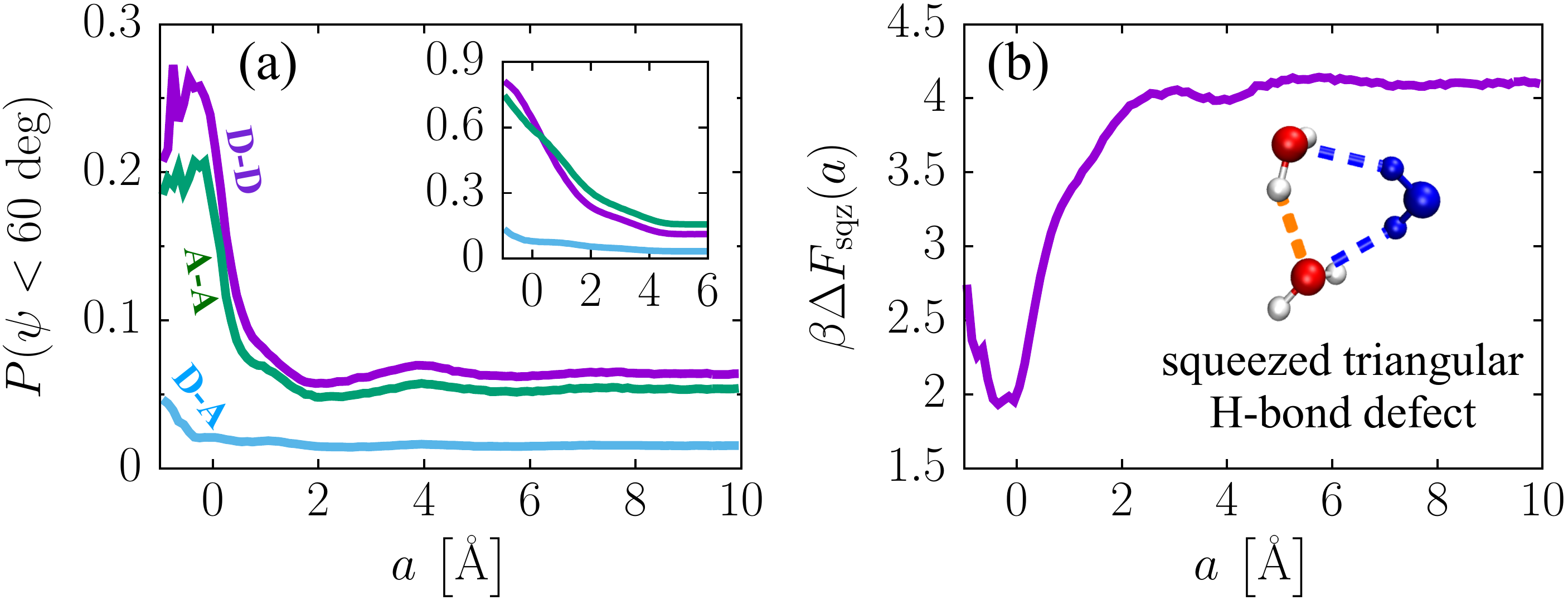}
\caption{(a) Probability of observing $\psi < 60\,\deg$ at varied interfacial depth from the atomistic simulation and our model (inset). Labels indicating the types of bond pairs are given in the same colors of lines. (b) Interfacial profile of free energy change associated with the distortion into a squeezed triangular hydrogen bond defect. Plot shown above is for the case including two donor bonds.}
\label{fig:4}
\end{figure}

We modify our empirical description of hydrogen bonding based on insight generated from analyzing the microscopic hydrogen bonding properties from atomistic simulation.
We quantify the hydrogen bond geometry of individual molecules by specifying the \emph{inter-bond} angle, $\psi$, between pairs of hydrogen bonds. 
As Fig.~\ref{fig:3}(a) illustrates, there are six such angles for a molecule with four unique hydrogen bond partners.
The probability distribution for $\psi$, $P(\psi)$, computed from atomistic simulation is plotted in Fig.~\ref{fig:3}(b). 
This plot highlights that the interfacial environment features highly distorted non-ideal hydrogen bond geometries that depend on the directionality (\emph{i.e.}, donor or acceptor) of the two adjacent hydrogen bonds.
Specifically interfacial inter-bond angles are narrowed relative to that of the bulk, and this narrowing is especially significant between bonds with like directionality.
Furthermore, as illustrated in Fig.~\ref{fig:4}(a), we observe a significant increase in the relative fraction of highly distorted hydrogen bond configurations within the first $2 \,\mathrm{\AA}$ of the interfacial region.

We quantify the energetic properties of these highly distorted hydrogen bonds by analyzing atomistic simulation data.
As Fig.~\ref{fig:3}(c) illustrates, the average direct interaction energy between a tagged molecule and one of its hydrogen bond partners is significantly weakened when the bond is distorted away from its preferred tetrahedral geometry.
Despite this weakening, however, we observe that highly distorted hydrogen bond configurations can be stabilized by favorable interactions between hydrogen bond partners.
As Fig.~\ref{fig:3}(d) illustrates, these interactions become particularly favorable when $\psi \approx 60\,\deg$, where the hydrogen bond partners are separated by approximately $d_\mathrm{HB}$, and thus well situated to form a hydrogen bond.
The resulting structure, a triangular three-body hydrogen bond defect, is unfavorable in the bulk liquid, however, at the interface this defect structure is stabilized by an increased availability of broken hydrogen bonds \cite{Geissler2013}, which due to the presence of the liquid phase boundary, cannot all be satisfied without significant distortions in hydrogen bond geometries.

Three-body interactions have been found to be an important element of water's molecular structure both in the bulk and at the interface \cite{Stillinger1973,Xantheas2000,Kumar2008,Medders2016}.
The three-body defects that we have identified here are structurally different from those that have been found to facilitate the orientational relaxation dynamics within the bulk liquid~\cite{Laage2006} and are uniquely stabilized at the interface.
We quantify the stabilizing effect of the interface by computing the relative free energy for squeezed triangular hydrogen bond defects, $\beta\Delta F_\text{sqz}(a) = -\ln \left[P_\text{sqz}(a)/(1-P_\text{sqz}(a))\right]$, where $P_\text{sqz}(a)$ denotes the probability to observe a molecule with position $a$ that is part of such a defect \cite{SM}. As Fig.~\ref{fig:4}(b) illustrates, these defects are more stable at the interface than in the bulk liquid by about $2 \,k_B T$. 

We incorporate the effects of these three-body interactions into our model by including angle dependence in the empirical hydrogen bond potential.
We accomplish this by allowing the angles of the hydrogen bonds to fluctuate, subject to the following expression, 
\begin{eqnarray}
E(\vec{\kappa},a, \lbrace n_k \rbrace&&) = \sum_{i=1}^4 \tilde{u}_\alpha(\phi_i)n_i \nonumber\\ &&+ \sum_{i<j} \left[\tilde{v}_{\alpha \gamma}(\psi_{ij}) - \lambda_{\alpha \gamma} (a_i, a_j, \psi_{ij})\right] n_i n_j,
\label{eq:energy2}
\end{eqnarray}
where $\phi_i$ denotes the angle of deviation of $\bp_i$ from its ideal tetrahedral orientation and $\psi_{ij}$ is the angle between $\bp_i$ and $\bp_j$ (see Fig.~\ref{fig:3}(a)).
In this expression the direct hydrogen bond energy, $\tilde{u}_\alpha(\phi)$, is described separately for donor and acceptor bonds, as denoted by the subscript $\alpha$. 
Similarly, the effects of three-body interactions are described by $\tilde{v}_{\alpha \gamma}(\psi)$, which depends on the directionality of the bonds involved in $\psi$, denoted by $\alpha$ and $\gamma$.
The function $\lambda_{\alpha \gamma}(a_i,a_j,\psi_{ij})$ is designed to attenuate the effects of three-body interactions based on the availability of broken hydrogen bonds at $a_i$ or $a_j$.
Here, we assign $\tilde{u}_\alpha$ and $\tilde{v}_{\alpha \gamma}$ to reflect the atomistic simulation data plotted in Fig.~\ref{fig:3}(c) and Fig.~\ref{fig:3}(d), respectively.
We refer to this variation of our model as the \textit{three-body fluctuation model}.
The details of this model variation are described in the Supplemental Material \cite{SM}.

The three-body fluctuation model exhibits interface-specific non-ideal hydrogen bond structure that is similar to that observed in atomistic simulation.
As the inset of Fig.~\ref{fig:4}(a) illustrates, this includes an interfacial enhancement of donor-donor and acceptor-acceptor angles of $\psi < 60\,\deg$, similar to that observed in atomistic simulation.
We find that including the effects of three-body hydrogen bond defects significantly improves the ability of the model to accurately describe interfacial polarization and polarizability, as illustrated in Fig.~\ref{fig:2}.
These effects do not, however, completely account for the interfacial variations in $\langle (\delta \mu_{\np}(a))^2 \rangle$, which suggests that interfacial polarizability also includes contributions from other microscopic effects, such as higher-order many-body effects.

The characteristics of interfacial molecular structure as derived from molecular dynamics simulations depend somewhat on the identity of the water force field. 
We illustrate some of these differences in the Supplemental Material~\cite{SM}.
We observe that different force fields exhibit similar trends in dipolar polarization and polarizability, but that they differ in their quantitative characteristics. 
These differences reflect subtle variations in force field structure that are not easily captured within the constraints of our mean field model.
We discuss this issue in more detail within the Supplemental Material~\cite{SM}.

To conclude, we summarize two fundamental aspects of interfacial molecular structure that have been revealed by our mean field model of the liquid water-vapor interface.
First, we observed that the depth dependent variations in molecular orientations at the interface are determined by the molecular geometry of hydrogen bonding and how it conforms to the anisotropic features of the interfacial density field.
We demonstrated that ideal bulk-like hydrogen bond geometry was sufficient to explain most of the interfacial variations in molecular orientations. 
Second, our model revealed that interfacial variations in molecular orientational polarization and polarizability arise due to distorted non-tetrahedral hydrogen bond structures.
We showed that the interface features squeezed triangular hydrogen bond defects that contribute significantly to determining interfacial dielectric properties.


We thank David Chandler, Liang Shi, and Rick C. Remsing for the useful discussions. This work was supported by the National Science Foundation under CHE-1654415 and also partially (SS) by the Kwanjeong Educational Foundation in Korea.



%

\pagebreak
\widetext
\begin{center}
\textbf{\large Supplemental Material: ``Three-body hydrogen bond defects contribute significantly to the dielectric properties of the liquid water-vapor interface"}
\end{center}
\setcounter{equation}{0}
\setcounter{figure}{0}
\setcounter{table}{0}
\setcounter{page}{1}
\makeatletter
\renewcommand{\theequation}{S\arabic{equation}}
\renewcommand{\thefigure}{S\arabic{figure}}
\renewcommand{\bibnumfmt}[1]{[S#1]}
\renewcommand{\citenumfont}[1]{S#1}

\section{Interfacial Structures of Different Classical Water Force Fields}
\begin{figure}[h!]
\centering
\includegraphics[width = 4.0 in]{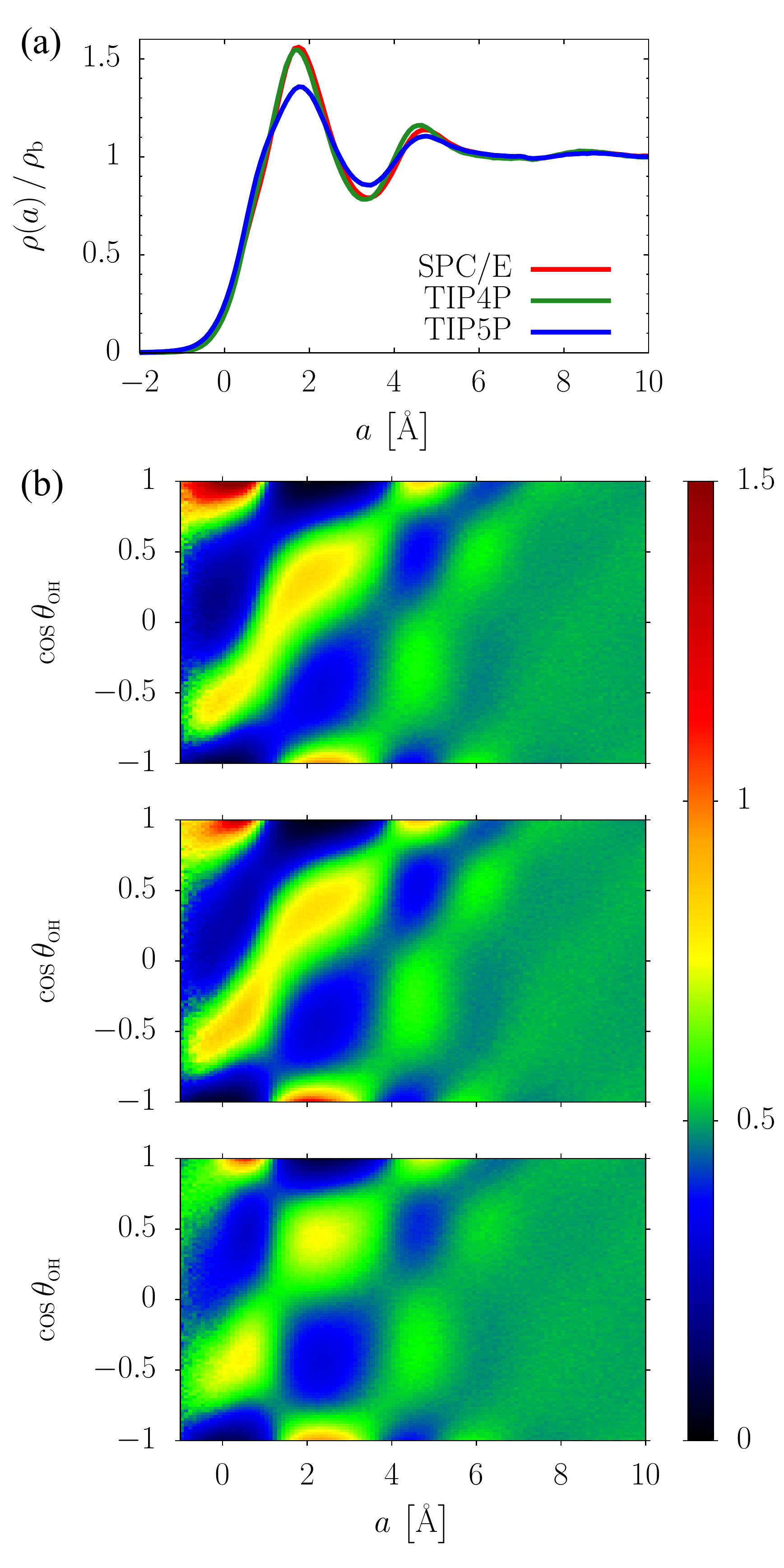}
\caption{(a) Density profiles computed from the atomistic simulations of SPC/E, TIP4P, and TIP5P. They are normalized by the bulk density, $\rho_\mathrm{b}$. (b) Orientational distributions, $P(\cos\theta_\text{OH}|a)$, for SPC/E (top), TIP4P (middle), and TIP5P (bottom). Color shading indicates the probability density.}
\label{fig:S1}
\end{figure}

To compare the interfacial structures of classical water force fields, we simulated the liquid-vapor interfaces of SPC/E \cite{Berendsen:1987gc}, TIP4P/2005 \cite{abascal2005general}, and TIP5P \cite{Mahoney2000_SM} waters in NVT ensembles. Each system has 1944 water molecules in a slab geometry with dimensions of 5.0 nm $\times$ 5.0 nm $\times$ 3.0 nm. The system was equilibrated at $T=298 \,\text{K}$ and Particle Mesh Ewald was used to handle the long-range part of electrostatic interactions. The SHAKE and SETTLE algorithms were used to constrain the geometry of water. The simulations were performed using the LAMMPS package \cite{Plimpton:1995fc} for SPC/E and TIP4P and the GROMACS package \cite{pronk2013gromacs} for TIP5P. Following the same procedure in Ref. \onlinecite{Willard:2010da}, instantaneous liquid interface was constructed for each configuration of the generated statistics. Density profile, $\rho(a)$, and orientational distribution, $P(\cos\theta_1,\cos\theta_2 | a)$, were computed according to Eqs. (7) and (11) in Ref. \onlinecite{Willard:2010da}. Figure \ref{fig:S1} shows the density profiles along with the reduced orientational distributions as defined in Eq. (5) of the main text. Both $\rho(a)$ and $P(\cos\theta_\text{OH}|a)$ exhibit the qualitatively same structural characteristics for three different force fields of water. 

\begin{figure}[h!]
\centering
\includegraphics[width = 6.4 in]{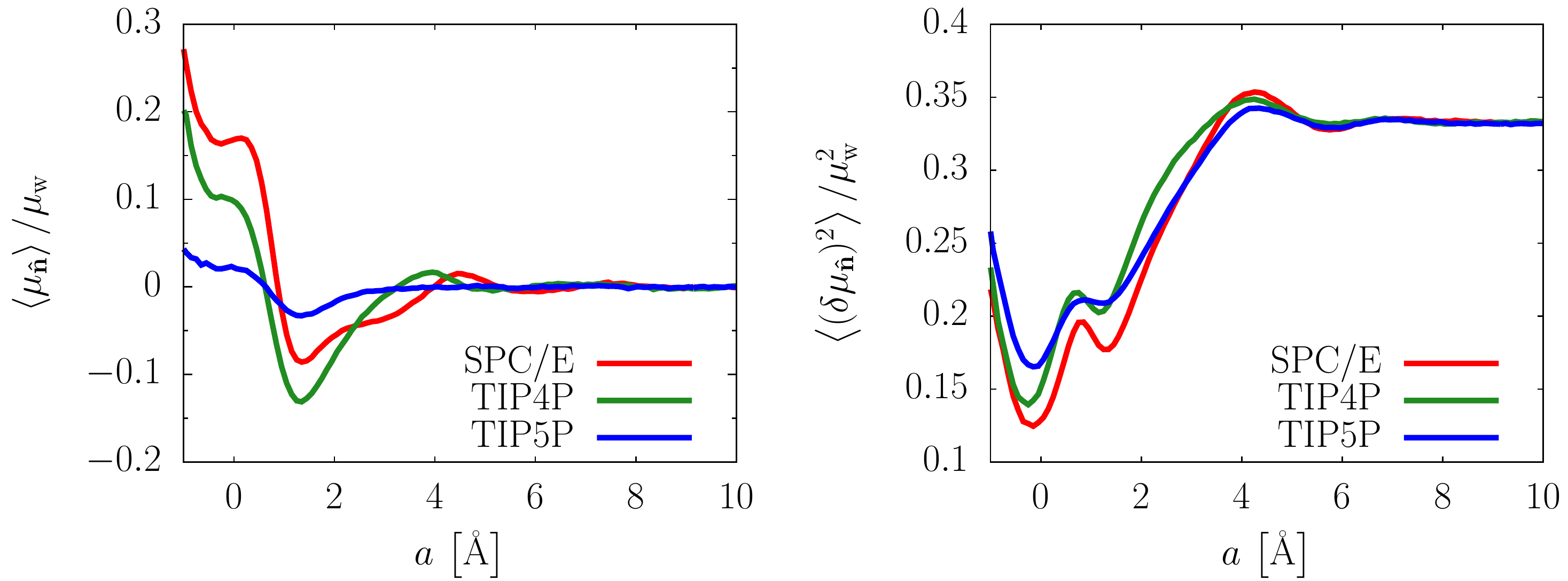}
\caption{Interfacial polarization, $\langle \mu_{\np}(a) \rangle$, and polarizability, $\langle (\delta \mu_{\np}(a) )^2 \rangle$, computed from the atomistic simulations of SPC/E, TIP4P, and TIP5P.}
\label{fig:S2}
\end{figure}

The orientational polarization, $\langle \mu_{\np}(a) \rangle$, and polarizability, $\langle (\delta \mu_{\np}(a) )^2 \rangle$, were computed according to Eqs.~(6) and (7) of the main text, and their interfacial profiles are plotted in Fig. \ref{fig:S2}. The plots show that these interfacial properties also share the qualitatively same feature among three different force fields of water. It is notable that while there exists some quantitative difference in the scale of the polarization, the polarizability shows the almost identical trend across the force fields.

\section{Optimization for the Hydrogen Bond Energy Parameter}
To determine the optimal model parameter, $\epsilon_\text{w}$, we compute the Kullback-Leibler divergence \cite{Kullback1951},
\begin{equation}
\Gamma(\epsilon_\text{w}) = \int d a \int d (\cos\theta_\text{OH}) \, P_\text{ref} (\cos\theta_\text{OH} | a) \ln \left[ \frac{ P_\text{ref} (\cos\theta_\text{OH} | a)}{P (\cos\theta_\text{OH} | a,\epsilon_\text{w})} \right],
\tag{S.1}
\label{gamma}
\end{equation}
where $P_\text{ref}(\cos\theta_\text{OH} | a)$ and $P (\cos\theta_\text{OH} | a,\epsilon_\text{w})$ are the reduced orientational distributions obtained from atomistic simulation and our mean-field model, respectively. This quantity measures how far the probability distribution of our model deviates from the reference given $\epsilon_\text{w}$. As indicated above, it becomes a function of $\epsilon_\text{w}$ and thus we choose the parameter that minimizes the \emph{fitness} function, 
\begin{equation}
\epsilon_\text{w}^* = \argmin_{\epsilon_\text{w}}\{\Gamma(\epsilon_\text{w})\},
\tag{S.2}
\end{equation}
as the effective hydrogen bond energy.


\section{Fluctuations in hydrogen bond geometry}
We quantify the distortions in hydrogen bond geometry and the associated energetics by analyzing the atomistic simulation results as follows.
Notably, here relatively simple algorithms for quantifying various aspects of molecular geometry translate into complicated mathematical expressions. 
Let $\mathbf{v}_{k}^{(i)} = \rp_k^{(i)} - \rp_\text{O}^{(i)}$, where $\rp_\text{O}^{(i)}$ is the position of the oxygen of the $i$th water molecule and $\rp_k^{(i)}$ is the position of the $k$th bonding site on it ($k=1,2$ indicate the hydrogens and $k=3,4$ indicate the lone pairs for TIP5P water). 
Then $\mathbf{v}_{k}^{(i)}$ represents the direction of ideal hydrogen bonding coordination through the $k$th site of the $i$th molecule.
For the $j$th molecule neighboring the $i$th, its deviation angle from the ideal coordination to the $i$th molecule is given by
\begin{equation}
\phi_j^{(i)} = \min_k \left\{\cos^{-1}\!\left( \frac{\mathbf{v}_{k}^{(i)} \cdot \bp_j^{(i)} }{\left|\mathbf{v}_{k}^{(i)}\right| \left|\bp_j^{(i)} \right|}\right) \right\},
\label{angle_phi}
\tag{S.3}
\end{equation}
where $\bp_j^{(i)} = \rp_\text{O}^{(j)}- \rp_\text{O}^{(i)}$ represents the hydrogen bond vector of the $i$th molecule to the $j$th. 
The corresponding index,
\begin{equation}
y_j^{(i)} = \argmin_k \left\{\cos^{-1}\!\left( \frac{\mathbf{v}_{k}^{(i)} \cdot \bp_j^{(i)} }{\left|\mathbf{v}_{k}^{(i)}\right| \left|\bp_j^{(i)} \right|} \right) \right\},
\tag{S.4}
\label{b_label}
\end{equation}
indicates which ideal boding direction $\bp_j^{(i)}$ is distorted from and thus whether it belongs to donor or acceptor hydrogen bond. 
The inter-bond angle between the $j$th and $k$th molecules with respect to the $i$th is given by
\begin{equation}
\psi_{jk}^{(i)} = \cos^{-1}\!\left( \frac{\bp_j^{(i)} \cdot \bp_k^{(i)} }{\left|\bp_j^{(i)}\right| \left|\bp_k^{(i)}\right|} \right).
\tag{S.5}
\label{angle_psi}
\end{equation}
Then the probability distribution of inter-bond angle at given distance $a$ is computed as,
\begin{equation}
P_{\alpha\gamma}(\psi|a) = \frac{\displaystyle \left< \sum_i \sum_{j \neq i} \sum_{k \neq i > j} \delta(\psi_{jk}^{(i)} - \psi)\delta(a^{(i)} - a) \Theta_\text{hyd}\!\left(\bp_j^{(i)} \right) \Theta_\text{hyd}\!\left(\bp_k^{(i)} \right) \Phi_\alpha \!\left(y_j^{(i)} \right) \Phi_\gamma \!\left(y_k^{(i)} \right) \right>}{\sin\psi\displaystyle\left< \sum_i \sum_{j \neq i} \sum_{k \neq i > j} \delta(a^{(i)} - a)\Theta_\text{hyd}\!\left(\bp_j^{(i)} \right) \Theta_\text{hyd}\!\left(\bp_k^{(i)} \right) \Phi_\alpha \!\left(y_j^{(i)} \right) \Phi_\gamma \!\left(y_k^{(i)} \right)\right>}~,
\tag{S.6}
\end{equation}
where $a^{(i)}$ is the interfacial depth of the $i$th molecule,
\begin{equation}
\Theta_\text{hyd}(\rp) = H(|\rp| - 2.4 \,\mathrm{\AA})H(3.2 \,\mathrm{\AA} - |\rp|)
\tag{S.7}
\end{equation} 
selects the molecules only in the first hydration shell of the $i$th molecule using the Heaviside step function, $H(x)$, and 
\begin{equation}
\Phi_\alpha \left(x \right) = \left\{ \begin{array}{ll}
		H\left(2.5 - x \right), & \quad   \textrm{if $\alpha$ is Donor}, \\
          	H\left(x - 2.5 \right), & \quad \textrm{if $\alpha$ is Acceptor}, \end{array} \right.
\tag{S.8}	
\end{equation}
selects the neighboring molecule of specific bond type, $\alpha$. Here the geometric factor, $\sin\psi$, corrects the bias coming from the variation of solid angle. 
Fig. 3(b) of the main text shows the plots of $P_{\alpha\gamma}(\psi|a)$ with $a = 10 \,\mathrm{\AA}$ and $a = 0 \,\mathrm{\AA}$ for the bulk and interface respectively ($0.1 \,\mathrm{\AA}$ was used for the binning width of histogram).
For the plots in Fig. 4(a) of the main text, the distribution is integrated such that $P_{\alpha\gamma}(\psi < 60^{\circ}|a) = \int_0^{60^{\circ}} P_{\alpha\gamma}(\psi|a)\sin\psi d\psi$.

Computing $P_\text{sqz}(a)$ needs to specify the certain type of defect among the configurations of $\psi < 60 
\,\text{deg}$. There is the other type of defect than the squeezed triangular one, which is known as the intermediate of the water reorientation \cite{Laage2006_SM}. 
This type of defect has bifurcated hydrogen bonds through one site of the molecule such that $y_j^{(i)} = y_k^{(i)}$. 
By excluding such cases, we can compute the probability to observe a squeezed configuration of two donor bonds as,
\begin{equation}
P_\text{sqz,DD}(a) = \frac{\displaystyle \left< \sum_i \sum_{j \neq i} \sum_{k \neq i > j} H\!\left(60^{\circ} - \psi_{jk}^{(i)} \right)\delta(a^{(i)} - a) \Theta_\text{hyd}\!\left(\bp_j^{(i)} \right) \Theta_\text{hyd}\!\left(\bp_k^{(i)} \right) \delta \!\left(y_j^{(i)}y_k^{(i)} - 2 \right) \right>}{\displaystyle\left< \sum_i \sum_{j \neq i} \sum_{k \neq i > j} \delta(a^{(i)} - a)\Theta_\text{hyd}\!\left(\bp_j^{(i)} \right) \Theta_\text{hyd}\!\left(\bp_k^{(i)} \right) \Phi_\text{D} \!\left(y_j^{(i)} \right) \Phi_\text{D} \!\left(y_k^{(i)} \right)\right>}~.
\tag{S.9}
\end{equation}

The average direct interaction energy for a hydrogen bond pair is computed in the bulk phase as a function of the deviation angle, $\phi$, such that
\begin{equation}
u_\alpha (\phi) = \frac{\displaystyle \left< \sum_i \sum_{j \neq i} U_{ij} \,\delta(\phi_j^{(i)} - \phi ) H\!\left(a^{(i)} - a_b\right)\Theta_\text{hyd}\!\left(\bp_j^{(i)} \right) \Phi_\alpha \!\left(y_j^{(i)} \right) \right>}{\displaystyle \left< \sum_i \sum_{j \neq i} \delta(\phi_j^{(i)} - \phi ) H\!\left(a^{(i)} - a_b\right) \Theta_\text{hyd}\!\left(\bp_j^{(i)} \right) \Phi_\alpha \!\left(y_j^{(i)} \right) \right>}~,
\tag{S.10}
\end{equation}
where $U_{ij}$ is the pair potential energy between the $i$th and $j$th molecules and $a_b = 10 \,\mathrm{\AA}$. The plots are given in Fig. 3(c) of the main text, normalized by the average bulk hydrogen bond energy where we used $E_\text{HB} = 9.0 \,k_B T$ (see the next section below for more details on this).
Similarly, the average direct interaction energy between two neighbors of a tagged molecule is computed as,
\begin{equation}
v_{\alpha\gamma} (\psi) = \frac{\displaystyle \left< \sum_i \sum_{j \neq i} \sum_{k \neq i > j} U_{jk} \,\delta( \psi_{jk}^{(i)} - \psi ) H\!\left(a^{(i)} - a_b\right) \Theta_\text{hyd}\!\left(\bp_j^{(i)} \right) \Theta_\text{hyd}\!\left(\bp_k^{(i)} \right) \Phi_\alpha \!\left(y_j^{(i)} \right) \Phi_\gamma \!\left(y_k^{(i)} \right) \right>}{\displaystyle \left< \sum_i \sum_{j \neq i}\sum_{k \neq i > j}  \delta (\psi_{jk}^{(i)} - \psi ) H\!\left(a^{(i)} - a_b\right) \Theta_\text{hyd}\!\left(\bp_j^{(i)} \right) \Theta_\text{hyd}\!\left(\bp_k^{(i)} \right) \Phi_\alpha \!\left(y_j^{(i)} \right) \Phi_\gamma \!\left(y_k^{(i)} \right) \right>}~.
\tag{S.11}
\end{equation}

\section{Implementation of the Three-body Fluctuation model} 
Following the notations used in the previous section, let $\{\mathbf{\hat{v}}_1, \mathbf{\hat{v}}_2, \mathbf{\hat{v}}_3, \mathbf{\hat{v}}_4\}$ be the unit vectors of ideal hydrogen bonding directions through the hydrogens and lone pairs of a probe water molecule of given orientation $\vec{\kappa}$.
We sample the hydrogen bond vectors, $\{\bp_1, \bp_2, \bp_3, \bp_4 \}$, each of which is within a certain solid angle around $\mathbf{\hat{v}}_i$ such that $\bp_i \cdot \mathbf{\hat{v}}_i = |\bp_i| \cos\phi_i$ where the value for $\cos\phi_i$ is drawn from the uniform random distribution of [$\cos 70^{\circ}$, 1]. 
Each hydrogen bond vector is assigned a directionality of either donor or acceptor based on the proximity to the bonding sites (see Eq.~(\ref{b_label})). 
The six inter-bond angles, $\psi_{ij}$, are calculated based on Eq.~(\ref{angle_psi}). 
If the bond vectors are too close with one another, \emph{i.e.} $\psi_{ij} < 44^{\circ }$, they are not taken into account for computing $P(\vec{\kappa}|a)$ since there is almost no statistics below that in the atomistic simulation. 
Hence $P(\vec{\kappa}|a)$ is computed as,
\begin{equation}
P(\vec{\kappa} \vert a) = \int \prod_{i=1}^{4} \left[d\bp_i \delta (|\bp_i| - d_\text{HB})H\!\left(\frac{\bp_i}{|\bp_i|} \cdot \mathbf{\hat{v}}_i - \cos 70^{\circ} \right)  \prod_{j > i} H\!\left(\cos44^{\circ} - \frac{\bp_i \cdot \bp_j}{|\bp_i||\bp_j|} \right)\right]  \frac{\left \langle e^{-\beta E(\vec{\kappa},a,\{n_k\})} \right\rangle_\mathrm{b}}{Z(a)},
\tag{S.12}
\label{eq:dist1}
\end{equation}
where $E(\vec{\kappa},a,\{n_k\})$ follows the Eq.~(8) of the main text. 
Here we impose the same constraint on the length of hydrogen bond vectors as that in the rigid tetrahedral model. 
Implementing the fluctuations in $|\bp_i|$ provokes more details about the energetics, $\tilde{u}_\alpha$ and $\tilde{v}_{\alpha\gamma}$, such as their dependence on both lengths and angles of the hydrogen bond vectors, which we have not detailed so far in this model. 

\begin{figure}[t!]
\centering
\includegraphics[width = 3.5 in]{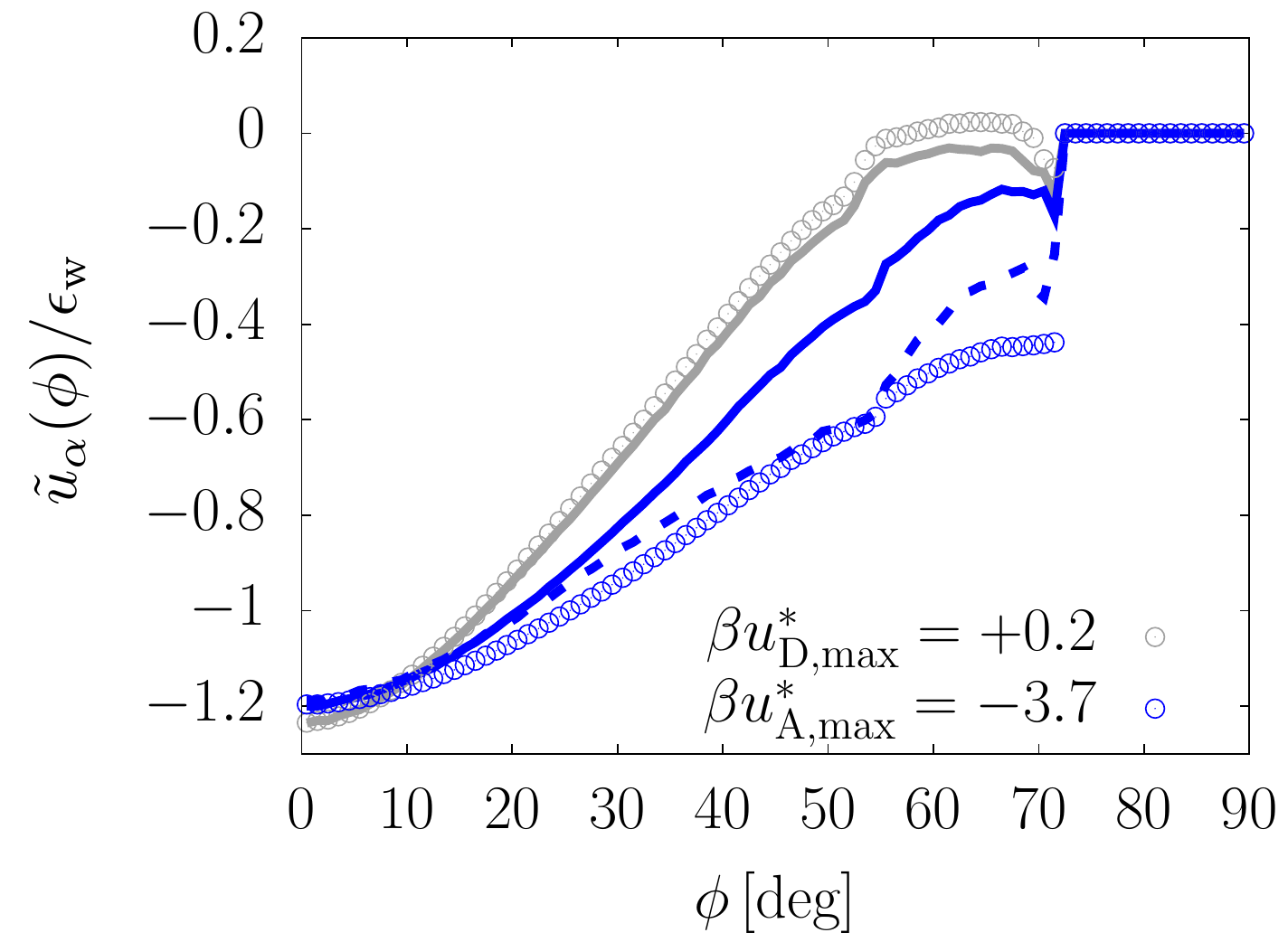}
\caption{$\tilde{u}_\alpha(\phi)$ compared with $u_\alpha(\phi)$ in the same scale. Solid lines are the simulation data of $u_\alpha(\phi)/E_\text{HB}$, as originally shown in Fig.~3(c) of the main text, and circles indicate the rescaled energy functions that are optimized for the accurate trend of $\langle \mu_{\np}(a) \rangle$. Gray and blue color correspond to $\alpha = \mathrm{D}$ and $\alpha = \mathrm{A}$, respectively. The blue dashed line corresponds to $u_\text{A}(\phi|a=1\,\mathrm{\AA})$, \emph{i.e.}, the average direct interaction energy for a hydrogen bond pair at $a = 1\,\mathrm{\AA}$.}
\label{fig:S3}
\end{figure}

The energy functions, $\tilde{u}_\alpha(\phi)$ and $\tilde{v}_{\alpha\gamma}(\psi)$, are rescaled from the atomistic simulation data of $u_\alpha(\phi)$ and $v_{\alpha\gamma}(\psi)$. 
Additionally, we parametrize $\tilde{u}_\alpha(\phi)$ by tuning the maximum value of $u_\alpha(\phi)$ such that
\begin{equation}
\tilde{u}_\alpha(\phi) = \left\{ \begin{array}{ll}
		\displaystyle\left\{ \big[u_\alpha(\phi) - u_\alpha(0) \big]\frac{u_\alpha^* - u_\alpha(0)}{u_{\alpha,\text{max}} - u_\alpha(0)} + u_\alpha(0) \right\}\frac{|\epsilon_\text{w}|}{E_\text{HB}}, & \quad\quad   \textrm{if $\phi \le \phi_c$},\;\;\quad\;\;\;\; \\\\
          	\displaystyle0, & \quad\quad \textrm{if $\phi > \phi_c$}, \end{array} \right.\\
\tag{S.13}
\label{eq:u_alpha}
\end{equation}
where $\phi_c = 72^{\circ}$, $u_{\alpha,\text{max}} = \max_{\phi < \phi_c}\left\{u_\alpha(\phi)\right\}$, and $u_\alpha^*$ is the parameter that sets the new maximum (in the original scale of $u_\alpha$). 
Here the factor of $|\epsilon_\text{w}|/E_\text{HB}$ rescales the functions in units of the effective hydrogen bond energy of our model.
We observed that the behavior of $\langle \mu_{\np}(a) \rangle$ is largely sensitive to $\tilde{u}_\alpha(\phi)$, and thus we optimized the parameters, $u_\alpha^*$ and $\epsilon_\text{w}$, for the result expected from atomistic simulations.
For the results presented in Fig. 2 of the main text, we used  $\epsilon_\text{w} = -5.0 \,k_B T$, $u_\text{D}^* = +0.2 \,k_B T$, and $u_\text{A}^* = -3.7 \,k_B T$. 
We found that the optimized $\tilde{u}_\text{A}(\phi)$ is quite different from $u_\text{A}(\phi)$ but more like the corresponding energy computed near the interface (see Fig. \ref{fig:S3}). 
For $\tilde{v}_{\alpha\gamma}(\psi)$, we simply take the values from the atomistic simulation data and rescale them in units of $\epsilon_\text{w}$, that is,
\begin{equation}
\tilde{v}_{\alpha\gamma} (\psi) = v_{\alpha\gamma}(\psi)\frac{|\epsilon_\text{w}|}{E_\text{HB}}~.
\tag{S.14}
\end{equation}

As given in Eq.~(8) of the main text, $\tilde{v}_{\alpha\gamma}(\psi)$ is combined with the auxiliary function, $\lambda(a_i,a_j,\psi_{ij})$, in order to accounts for the interface-specific stability of hydrogen bond defects.
This auxiliary function represents the energetic cost for the defects to pay based on the hydrogen bonding status of the $i$th or $j$th hydrogen bond partner. 
Assuming that this penalty is imposed on the one that donates hydrogen, we describe this function in terms of the average number and energy of hydrogen bonds through donor sites, denoted by $N_\text{D} (a)$ and $E_\text{D} (a)$ respectively. 
Specifically it is given by,
\begin{equation}
\lambda_{\alpha\gamma} (a_i,a_j,\psi) = \left\{ \begin{array}{ll}
		\displaystyle \frac{N_\text{D} (\bar{a}_{\alpha\gamma})}{2}E_\text{D} (\bar{a}_{\alpha\gamma})\frac{|\epsilon_\text{w}|}{E_\text{HB}}, & \quad   \textrm{if $\psi \le \psi_{\alpha\gamma}$}, \\\\
		\displaystyle \left[ 1 - \frac{ v_{\alpha\gamma}(\psi) - v_{\alpha\gamma}( \psi_{\alpha\gamma})}{v_{\alpha\gamma}(\psi_c) - v_{\alpha\gamma}(\psi_{\alpha\gamma}) } \right] \lambda_{\alpha\gamma}(a_i,a_j,\psi_{\alpha\gamma}), & \quad   \textrm{if $\psi_{\alpha\gamma} < \psi < \psi_c$}, \\\\
          	\displaystyle 0, & \quad \textrm{if $\psi \ge \psi_c$}, \end{array} \right.
\tag{S.15}
\end{equation}
where $\psi_{\alpha\gamma} = \argmin_{\psi}\left\{v_{\alpha\gamma}(\psi)\right\}$, $\psi_c = \argmax_{\psi}\left\{v_\text{AA}(\psi)\right\}$, and
\begin{equation}
\bar{a}_{\alpha\gamma} (a_i,a_j) = \left\{ \begin{array}{ll}
		\displaystyle \min \{a_i, a_j \}, & \quad   \textrm{if $\alpha = \gamma$},\\\\
          	\displaystyle \Phi_\text{D}\!\left(y_i \right)a_i + \Phi_\text{D}\!\left(y_j \right)a_j, & \quad \textrm{if $\alpha \ne \gamma$}. \end{array} \right.
\tag{S.16}
\end{equation}
Here we let the penalty taken by the hydrogen bond partner located closer to the interface, but we make the exception for donor-acceptor bond pairs based on the typical structure of cyclic water trimer \cite{keutsch2003water}.
$N_\text{D}(a)$ and $E_\text{D}(a)$ are computed from atomistic simulation as, 
\begin{equation}
N_\text{D} (a) = \frac{\displaystyle \left< \sum_i \sum_{j \neq i} \delta(a^{(i)} - a) H\!\left(30^{\circ} - \phi_j^{(i)} \right) H\!\left(3.5 \,\mathrm{\AA} - \left\vert \bp_j^{(i)}\right\vert \right) \Phi_\text{D}\!\left(y_j^{(i)} \right) \right>}{\displaystyle \left< \sum_i \delta\!\left(a^{(i)} - a \right) \right>}~,
\tag{S.17}
\end{equation}
and
\begin{equation}
E_\text{D} (a) = \frac{\displaystyle \left< \sum_i \sum_{j \neq i} U_{ij} \, \delta(a^{(i)} - a) H\!\left(30^{\circ} - \phi_j^{(i)} \right) H\!\left(3.5 \,\mathrm{\AA} - \left\vert \bp_j^{(i)}\right\vert \right) \Phi_\text{D}\!\left(y_j^{(i)} \right) \right>}{\displaystyle \left< \sum_i \sum_{j \neq i} \delta(a^{(i)} - a) H\!\left(30^{\circ} - \phi_j^{(i)} \right) H\!\left(3.5 \,\mathrm{\AA} - \left\vert \bp_j^{(i)}\right\vert \right) \Phi_\text{D}\!\left(y_j^{(i)} \right) \right>}~,
\tag{S.18}
\end{equation}
where the definition of a good hydrogen bond follows the one by Luzar and Chandler \cite{luzar1996effect}.
\begin{figure}[t!]
\centering
\includegraphics[width = 6.4 in]{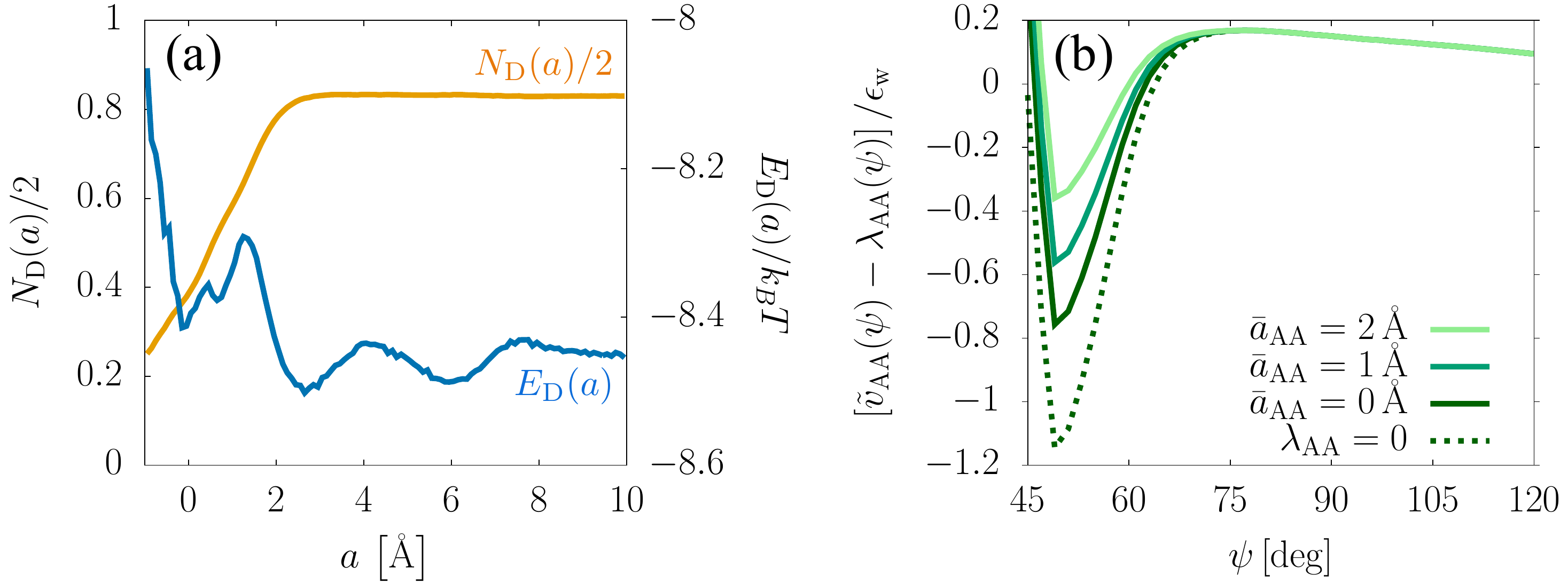}
\caption{(a) Interfacial profiles of average number and energy of hydrogen bonds through donor sites, rendered in orange and blue lines respectively. (b) Illustration of a three-body interaction term implemented in our model. Solid lines show the different energetic preferences for highly distorted configurations depending on the interfacial depth of hydrogen bond partner. Dashed line corresponds to $\tilde{v}_\text{AA}$ without the attenuation by $\lambda_\text{AA}$.}
\label{fig:S4}
\end{figure}
As illustrated in Fig.~\ref{fig:S4}, they change dramatically within the first $2\,\mathrm{\AA}$ of the interfacial region such that the effect of three-body interaction also becomes significant in that region. 
Here we take the value of $E_\text{HB}$ from $E_\text{D}(a)$ by setting $E_\text{HB} = |E_\text{D}(a_b)| = 8.45 \,k_B T = 20.9 \,\mathrm{kJ/mol}$ at $T = 298 \text{ K}$.

\begin{figure}[t!]
\centering
\includegraphics[width = 3.4 in]{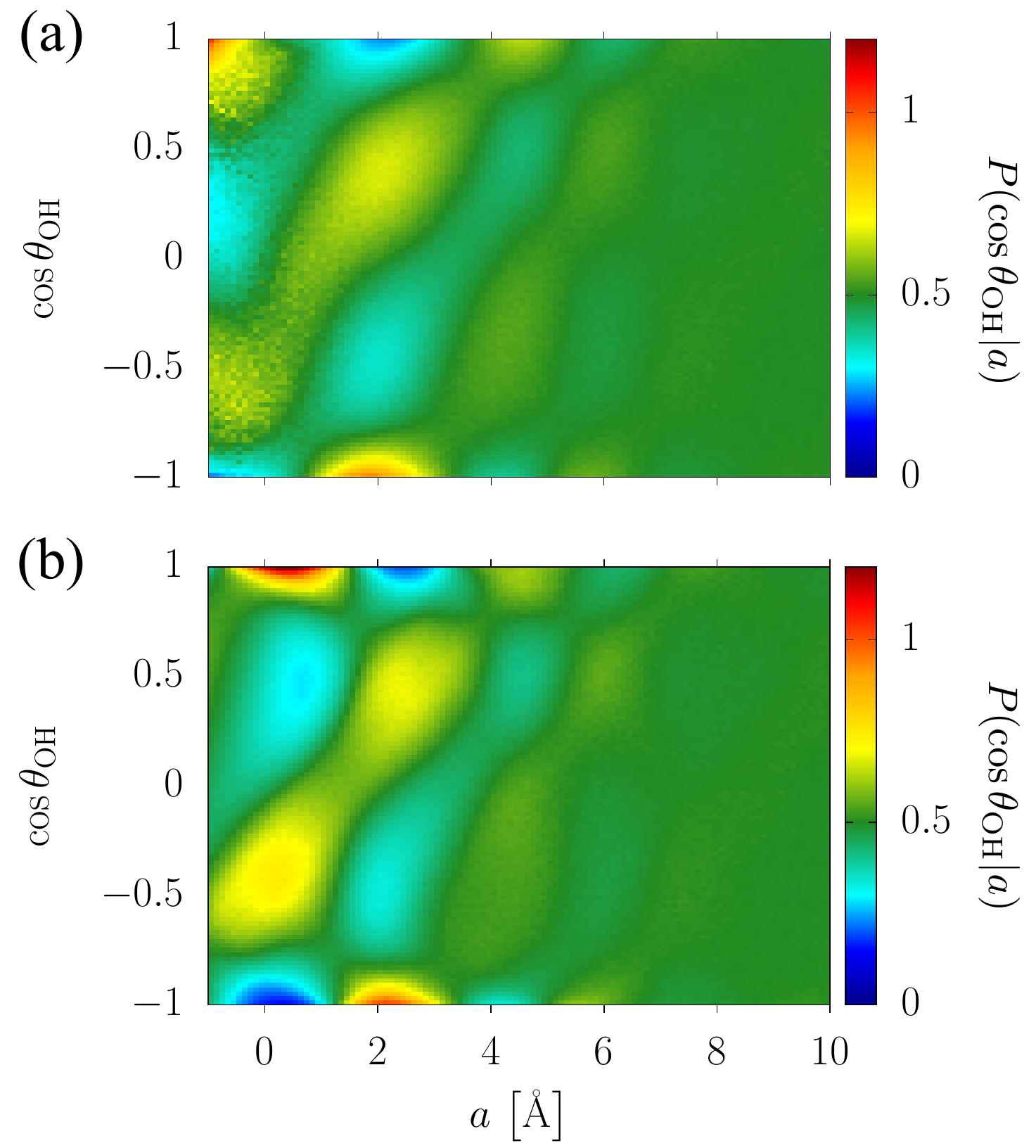}
\caption{Orientational distributions, $P(\cos\theta_\text{OH}|a)$, computed from (a) the three-body fluctuation model and (b) the rigid tetrahedral model for the TIP5P force field. Color shading indicates the probability density.}
\label{fig:S5}
\end{figure}

In order to evaluate $P(\vec{\kappa}|a)$, we obtain an approximate analytic expression for $\left< e^{-\beta E(\vec{\kappa},a,\{n_k\})} \right>_\text{b}$ in Eq.~(\ref{eq:dist1}).
Here we made the same assumption as that of the rigid tetrahedral model, such that $\langle n_i n_j \rangle \approx \langle n_i \rangle \langle n_j \rangle $ for $i \ne j$
\footnote{
This is definitely a rough approximation since it neglects the density correlation between hydrogen bond partners even in the squeezed configurations. However, more accurate treatment for the density correlation provokes again the distance dependence of the energy functions which we have not detailed so far in this model. 
}. 
Within this approximation, we can write
\begin{align*}
\left< e^{-\beta E(\vec{\kappa},a,\{n_k\})}\right>_\text{b} &= \prod_{i=1}^4 \langle n_i \rangle_\text{b} e^{-\beta \tilde{u}_\alpha(\phi_i)} \prod_{j > i}^{4} e^{-\beta  \tilde{v}_{\alpha\gamma}(\psi_{ij})} + \sum_{k=1}^4 \left[ 1 - \langle n_k \rangle_\text{b}\right] \prod_{i \neq k}^4 \langle n_i \rangle_\text{b} e^{-\beta \tilde{u}_\alpha(\phi_i)} \prod_{j\neq k > i}^4 e^{-\beta  \tilde{v}_{\alpha\gamma}(\psi_{ij})} \nonumber\\ &\quad+ \sum_{i = 1}^4\sum_{j > i}^4  \langle n_i \rangle_\text{b}\langle n_j \rangle_\text{b} \left[1 - \langle n_k \rangle_\text{b} \right] \left[1 - \langle n_l \rangle_\text{b} \right] e^{-\beta \left[\tilde{u}_\alpha(\phi_i) + \tilde{u}_\alpha(\phi_j) + \tilde{v}_{\alpha\gamma}(\psi_{ij}) \right]} \nonumber\\ &\quad\quad+ \sum_{k=1}^4 \langle n_k \rangle_\text{b} e^{-\beta \tilde{u}_\alpha(\phi_k)} \prod_{i \neq k}^4 \left[1 - \langle n_i \rangle_\text{b}\right] + \prod_{i=1}^4 \left[1 - \langle n_i \rangle_\text{b}\right]~,\nonumber
\tag{S.19}
\end{align*}
where  $\langle n_i \rangle_\text{b} = P_\text{HB} (a_i) = \rho(a_i)/2\rho_b$ and the dummy indices, $k$ and $l$, in the third term are the numbers among $\{1,2,3,4\}$ such that $i \ne j \ne k \ne l$. 
The resulting reduced probability distribution, $P(\cos\theta_\text{OH}|a)$, is given in Fig.~\ref{fig:S5}.
Although its qualitative feature is still the same as the result from the rigid tetrahedral model, its details are closer to that from the atomistic simulation of TIP5P water (especially at $a < 3 \,\mathrm{\AA}$).

\section{Application of mean-field models to SPC/E force field}
\begin{figure}[t!]
\centering
\includegraphics[width = 3.0 in]{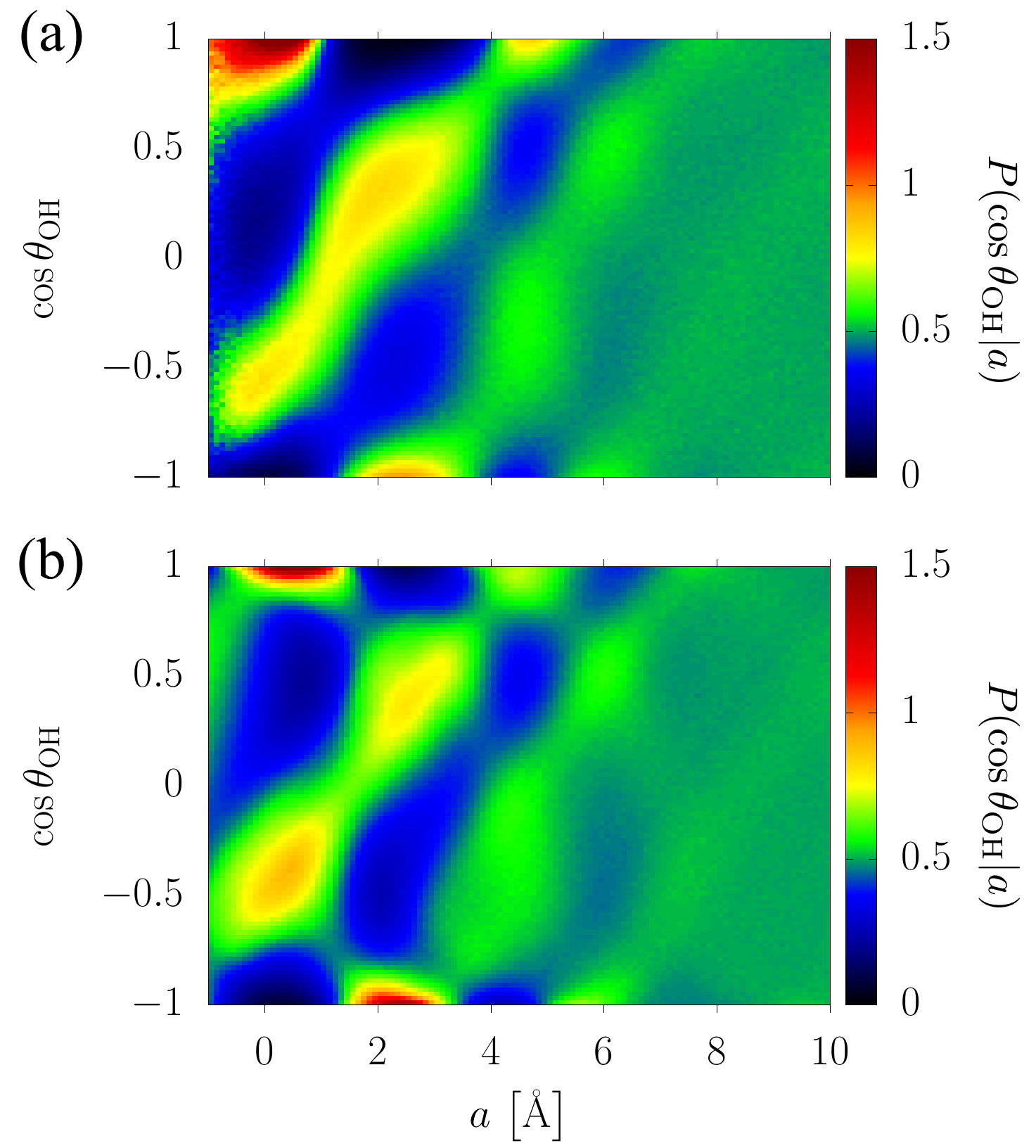}
\caption{$P(\cos\theta_\text{OH}|a)$ computed from (a) the atomistic simulation with SPC/E water and (b) the rigid tetrahedral model optimized for the SPC/E force field ($\epsilon_\text{w}^* = -1.8\,k_B T = -4.46 \text{ kJ/mol}$).}
\label{fig:S6}
\end{figure}

\begin{figure}[h!]
\centering
\includegraphics[width = 5.0 in]{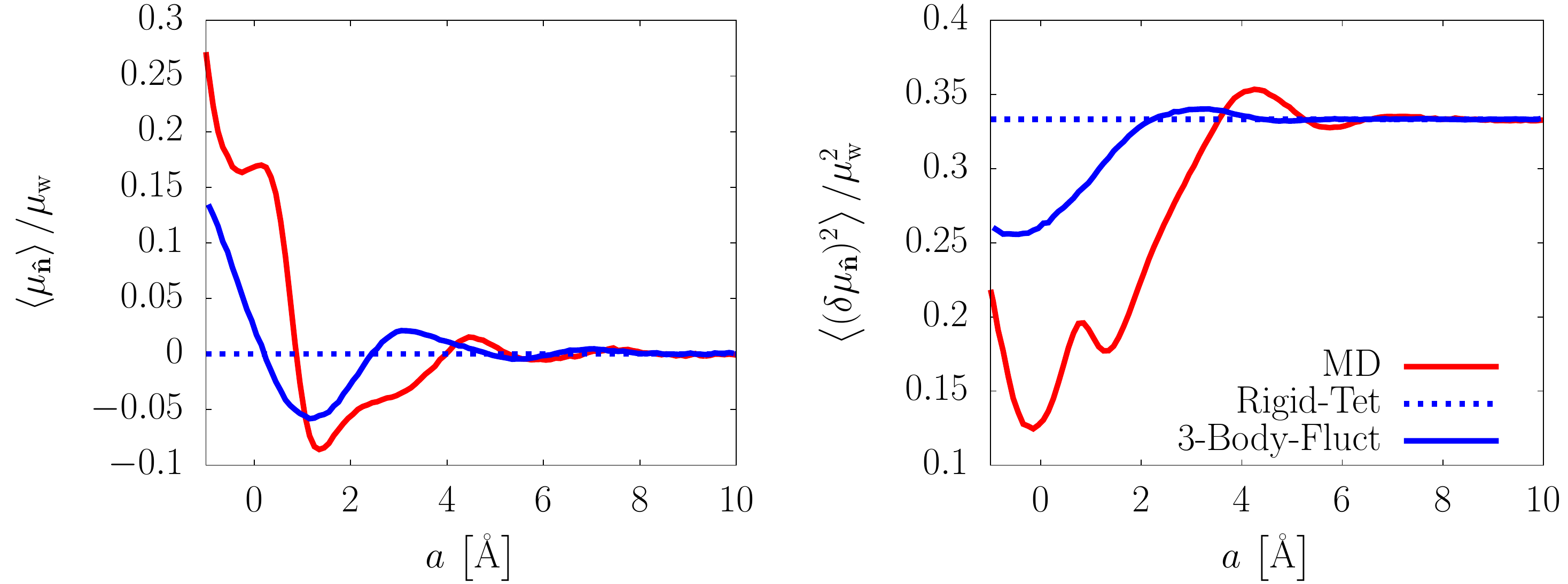}
\caption{Interfacial polarization, $\langle \mu_{\np}(a) \rangle$, and polarizability, $\langle (\delta \mu_{\np}(a) )^2 \rangle$, computed from the three-body fluctuation model parametrized for the SPC/E force field, in comparison to the molecular dynamics simulation results.}
\label{fig:S7}
\end{figure}

Despite sharing similar bulk hydrogen bonding structures, different classical water models can yield non-trivial differences in interfacial structure. This is highlighted to some extent in Figs.~\ref{fig:S1} and \ref{fig:S2}. As Fig.~\ref{fig:S2} illustrates, different water models exhibit similar trends in $\langle \mu_{\np}(a) \rangle$ and $\langle (\delta \mu_{\np}(a) )^2 \rangle$, but they differ in their quantitative characteristics. To evaluate the ability of our mean field model to capture these differences have also applied our model to the SPC/E force field. To do this, we have followed the same procedure described herein but using the data from molecular dynamics simulations of SPC/E water rather than TIP5P. For the SPC/E-parameterized mean field model we have found similarly good agreement in reproducing $P(\cos\theta_\text{OH}|a)$, as illustrated in Fig.~\ref{fig:S6}, but as Fig.~\ref{fig:S7}  illustrates, the agreement to $\langle \mu_{\np}(a) \rangle$ and $\langle (\delta \mu_{\np}(a) )^2 \rangle$, is not as strong for SPC/E as it is for TIP5P. 

We understand that the difference in the ability of our model to reproduce dipolar polarization/polarizability between SPC/E and TIP5P arises due to the differences in the geometric tendencies inherent to these force fields. The TIP5P force field is built upon a tetrahedral charge scaffold, so non-ideal hydrogen bond structures are more naturally described in terms of their deviations from this scaffold. On the other hand, SPC/E is built upon a triangular charge scaffold (albeit with a tetrahedral bond angle) so non-ideal hydrogen bond structures are less well represented in term of deviation from a tetrahedral scaffold. We speculate that we could improve quantitative accuracy of our model in reproducing SPC/E results by modifying the details of the underlying geometry of our mean field model.

\end{document}